\documentclass[12pt]{article}
\usepackage{amsfonts}
\usepackage[utf8]{inputenc}
\usepackage[english]{babel}
\usepackage{bm}
\usepackage{amsmath}
\usepackage{epsfig}
\usepackage{enumitem}

\setlist[enumerate]{%
	wide =0.5\parindent,
	listparindent=0pt%
}

\newcommand{\bmat}{\left(\begin{array}}
	\newcommand{\emat}{\end{array}\right)}

\def\gtrsim{\mathrel{\raise.3ex\hbox{$>$\kern-.75em\lower1ex\hbox{$\sim$}}}}

\def\ov{\overline}
\def\un{\underline}

\def\-{\hphantom{-}}
\def\ov{\overline}
\def\s2{\frac{1}{\sqrt2}}

\def\h{\hat}

\def\mg{m_{3/2}}
\def\mg2{m^2_{3/2}}

\def\Dsl{\,\raise.15ex\hbox{/}\mkern-13.5mu D} 

\def\be{\begin{equation}}
	\def\ee{\end{equation}}
\def\bea{\begin{eqnarray}}
	\def\eea{\end{eqnarray}}

\newcommand{\nn}{\nonumber}

\begin{document}
	
	
	\pagestyle{plain}

	\makeatletter
	\@addtoreset{equation}{section}
	\makeatother
	\renewcommand{\theequation}{\thesection.\arabic{equation}}
	\pagestyle{empty}
	\begin{center}
		\ \
		
		\vskip .5cm
		
		\LARGE{\LARGE\bf Double Field Theory with matter and its cosmological application}
\vskip 0.3cm

\large{Eric Lescano$^{1}$ and Nahuel Mir\'on--Granese$^{2,3,4}$ 
	\\[3mm]}

{\small  $^1$ Division of Theoretical Physics, Ruđer Bošković Institute\\ [.01 cm]}
{\small\it Bijenička 54, 10000 Zagreb, Croatia\\ [.2 cm]}

{\small  $^2$ Consejo Nacional de Investigaciones Científicas y Técnicas (CONICET)\\ [.01 cm]}
{\small\it Rivadavia 1917, Ciudad de Buenos Aires (1428), Argentina\\ [.2 cm]}

{\small  $^3$ Facultad de Ciencias Astronómicas y Geofísicas, Universidad Nacional de La Plata\\ [.01 cm]}
{\small\it Paseo del Bosque, La Plata (B1900FWA), Buenos Aires, Argentina\\ [.2 cm]}

{\small  $^4$ Departamento de Física, Facultad de Ciencias Exactas y Naturales,\\ Universidad de Buenos Aires\\ [.01 cm]}
{\small\it Ciudad Universitaria, Pabellón 1, Ciudad de Buenos Aires (1428), Argentina\\ [.2 cm]}

{\small \verb"elescano@irb.hr, nahuelmg@fcaglp.unlp.edu.ar"}\\[0.5cm]
\small{\bf Abstract} \\[0.5cm]\end{center}

The phase space formulation of Double Field Theory (DFT) indicates that statistical matter can be included in terms of (T--)duality multiplets. We propose the inclusion of a perfect fluid in the geometry of DFT through a generalized energy--momentum tensor written in terms of a DFT pressure, energy density and velocity. The latter is an $O(D,D)$ vector and satisfies two invariant constraints, in agreement with the on--shell constraints for the generalized momentum. We compute the conservation laws associated to the energy-momentum tensor considering general DFT backgrounds. Then we study cosmological backgrounds using a space--time split, and we find an expression for the DFT cosmological dynamics with the perfect fluid coupled. This proposal reproduces the equations of string cosmology upon parametrization of the DFT Einstein equations.
\newpage

\setcounter{page}{1}
\pagestyle{plain}
\renewcommand{\thefootnote}{\arabic{footnote}}
\setcounter{footnote}{0}

\tableofcontents
\newpage

\section{Introduction}
String Theory (ST) is a very good candidate to describe nature from first principles. One of the cornerstones of ST is duality, which is necessary to show the equivalence between their different formulations. Moreover, T--duality can be used as a guiding principle to construct a double geometry that describes strings in an $O(D,D)$ invariant way, $O(D,D)$ being an exact symmetry of ST. One possible framework to study the interplay between T--duality and the low energy limit of ST is Double Field Theory  \cite{Siegel,DFT,DFTkorea} (DFT)\footnote{For reviews see \cite{ReviewDFT1} and \cite{ReviewDFT2}.}. The main idea of DFT is to accomplish $O(D,D)$ as a symmetry of the effective (or supergravity) action. The fundamental dimension of $O(D,D)$ is $2D$ and therefore the dimensions of the target space must be doubled as $\hat X^{M}=(\tilde x_{\mu},x^{\mu})$ with $\mu=0,\,\dots,\,D-1$. 

The minimal formulation of DFT includes an invariant group (non--dynamical) metric, $\hat \eta_{\hat M \hat N}$, and a field content given by a generalized metric $\hat {\cal H}_{\hat M \hat N}(\hat X)$ and a generalized dilaton $\hat d(\hat X)$ which are  multiplets of the duality group. Standard diffeomorphisms and Abelian gauge transformations of the b-field are replaced by generalized diffeomorphisms in order to preserve the invariance of the group metric, i.e., $\delta_{\hat \xi} \hat \eta_{\hat M \hat N} = 0$. This formulation describes the universal NS-NS sector of the low energy limit of ST upon suitable parametrization of the fields and parameters and imposing the strong constraint, which ensures the closure of the generalized diffeomorphisms and removes the dependence on half of the coordinates. 

The construction of DFT is inspired in toroidal compactifications where T--duality transformations appear as a symmetry and the fundamental fields can be cast in multiplets of the duality group. These compactifications, in turn, are compatible with string cosmology \cite{Gasperini}, string gas cosmology \cite{SGC} and other scenarios with matter terms \cite{VafaTseytlin}. At the level of the worldsheet, this issue was addressed in \cite{90s} for cosmological vacuum solutions and in \cite{SFD,GasVene} considering matter contributions and their relation with T--duality. Moreover in \cite{VafaTseytlin} a duality covariant formulation was obtained for string cosmology. With these results, it is expected that matter fields can be coupled to the standard formulation of DFT, as it was recently stressed in \cite{HZbullet}.

When coupling matter fields to the vacuum DFT at the level of the action, the equations of motion derived from the variational principle can be recast into the form of a generalized Einstein equation \cite{ParkCosmo} \bea
\hat {\cal G}_{\hat M \hat N}=\mathcal{\h T}_{\h M \h N}\, ,\label{einsteindftintro}
\eea
where $\hat {\cal G}_{\hat M \hat N}$ is the symmetric DFT Einstein tensor and
\bea
\mathcal{\h T}_{\h M \h N} & = & \mathcal{\h H}_{\hat M \hat N}\left({\mathcal L}_m-\frac12\frac{\delta {\mathcal L}_m}{\delta \hat d}\right)-2\,\Big[\overline{P}_{\h M \h K} P_{\h N \h L}+\overline{P}_{\h N \h K} P_{\h M \h L}\Big]\left(\frac{\delta {\cal L}_m}{\delta {P}_{\h K \h L}}-\frac{\delta {\cal L}_m}{\delta \overline{P}_{\h K \h L}}\right)
\label{EMTDFT}
\eea
is the corresponding energy--momentum tensor with ${\cal L}_{m}$ the matter Lagrangian coupled to the double geometry and $P_{\hat M \hat N}$, $\bar P_{\hat M \hat N}$ the DFT projectors\footnote{We use the following conventions for the projectors: $\bar{P}_{\h M \h N}=\frac12(\hat \eta_{\h M \h N} + \h{\cal H}_{\h M \h N})$ and ${P}_{\h M \h N}=\frac12(\hat \eta_{\h M \h N} - \h{\cal H}_{\h M \h N})$. We do not include hats in the $O(D,D)$ projectors to present a light notation.}. Additionally the generalized Bianchi identities imply a vanishing divergence of the Einstein tensor, $\nabla^{\hat M} \mathcal{\h G}_{\h M \h N} = 0$, and in turn the on--shell condition dictates a conservation law for matter, namely
\bea
\nabla^{\hat M} \mathcal{\h T}_{\h M \h N} = 0 \, . 
\label{conserv}
\eea

The vacuum solutions for the DFT cosmological ansatz were studied in \cite{Wu} and further solutions with matter were explored in \cite{ParkCosmo,Branden}. Moreover, higher--derivative terms were included in \cite{HZbullet,Brandenmatter,alphapcosmo,Chouha} as well. Also in here we observe these works suggest that matter can be coupled to the standard construction of DFT. In turn having a well understood framework of matter coupled to DFT would allow to generalize the current computations of $\alpha'$ corrections using the results of \cite{gbdr}, where a systematic procedure to access to higher--derivative terms for vacuum DFT was introduced. 

If one wants to describe matter from a statistical approach, as a gas or fluid in the double geometry, it is possible to proceed from a double kinetic perspective \cite{phaseDFT} \footnote{See also \cite{VafaTseytlin, Branden} where a different procedure to include matter was implemented.}. Considering the phase space construction of DFT through the coordinates $(\hat X^{\hat M},\hat {\cal P}^{\hat M})$, with $\hat {\cal P}^{\hat M}$ being a generalized momentum vector, the analogous equation to (\ref{EMTDFT}) is given by
\bea
\hat{\cal T}^{\hat M \hat N}(\h X) = \int d^{2D}\h{\cal P}\,e^{-2\h d} \,
\h {\cal P}^{\hat M} \h{\cal P}^{\h N}  \,\h F  \, ,\label{emtintro}
\eea
where $\h F=\h F(\h X, \h {\cal P})$ is a generalized one--particle distribution function. In this double kinetic framework, the conservation law (\ref{conserv}) for the energy--momentum tensor (\ref{emtintro}) can be obtained from the transfer equations related to the generalized Boltzmann equation for the evolution of $\h F$.

\subsection{Main results}

While in General Relativity (GR) the explicit form of the energy--momentum tensor is well--known for several specific scenarios, e.g. the perfect fluid, this is not the case for DFT. The explicit form of the generalized distribution function is not known in any configuration since, in particular, the thermodynamics properties of the system and their equilibrium states are not fully understood. To avoid this last issue, in this work we introduce a proposal for the energy--momentum tensor of the perfect fluid in the geometry of DFT.

The ansatz is constructed taking a top--down perspective and proposing a generalization of the energy--momentum tensor for the perfect fluid in GR, adapting its structure in order to fulfill all the DFT constraints. Moreover, since DFT encodes the low energy limit of ST, this proposal must recover the fundamental equations of string cosmology. Although there exists a straightforward generalization of the definitions of the scalars (energy density and pressure), the construction of the generalized velocity of a point--particle in a particular orbit of the double phase space is more subtle. The generalized velocity for the particle in the double geometry is defined as
\bea
\hat U^{\hat M} = \frac{D\hat X^{\hat M}}{D\tau} \, ,\label{uintro}
\eea
where $D\h X^{\h M}$ is the DFT--compatible 1--form differential which transforms as \cite{Prk},
\bea
\delta_{\xi} (D \hat X^{\hat M}) = (\partial_{\hat N} \xi^{\hat M} - \partial^{\hat M} \xi_{\hat N})D \hat{X}^{\hat N} \, ,
\eea
and $\tau$ is an affine parameter in the double geometry, which reduces to the standard proper time upon parametrization. 

Both the generalized momentum $\hat {\mathcal P}^{\hat M}$ corresponding to the trajectory of a point--particle and its generalized velocity $\hat U^{\hat M}$ are vectors that indicate the same geometric direction in the tangent space and then they must be proportional. Choosing the proportionality constant is, in fact, defining an specific affine parameter $\tau$ in (\ref{uintro}). Indeed, we take the relation 
\bea
\hat {\mathcal P}^{\hat M}=m\,\hat U^{\hat M}=m\,D\hat X^{\hat M}/D\tau \, , 
\eea
where $m$ is the invariant mass. Since the generalized momentum satisfies a mass--shell condition and a strong constraint--like equation,
\bea
\label{constraintsintro0}
\h{\cal P}^{\h M} \h{\cal H}_{\h M \h N} \h{\cal P}^{\h N} & = & -m^2 \, , \\ \h{\cal P}^{\h M} \h{\eta}_{\h M \h N} \h{\cal P}^{\h N} & = & 0 \, ,
\label{constraintsintro1}
\eea
the generalized velocity inherits the following relations
\begin{eqnarray}
\hat U_{\hat M}  \h{\cal H}^{\hat M\hat N}\hat U_{\hat N}&=&-1\label{Uprop1intro}\\
\hat U_{\hat M}  {\hat \eta}^{\hat M\hat N} \hat U_{\hat N}&=&0 \, .
\label{Uprop2intro}
\end{eqnarray}   

The mass--shell condition for the particle in the double geometry (\ref{constraintsintro0}) is related to the mass operator and level matching condition \cite{ReviewDFT1}. The constraints (\ref{constraintsintro1}) and (\ref{Uprop2intro}) take rid off the half of the components of the generalized velocity/momentum, in agreement with the number of degrees of freedom of these objects. Furthermore, (\ref{Uprop1intro}) and (\ref{Uprop2intro}) allow us to consider a rest frame upon a splitting of the DFT coordinates. 

On the other hand, if we focus our attention on the structure of the energy--momentum tensor for the perfect fluid in GR, i.e.
\bea
T_{\mu \nu}=(p+e) u_{\mu} u_{\nu} + p g_{\mu \nu} \, ,\label{tmunugrintro} 
\eea
we can straightforwardly generalize the second term in (\ref{tmunugrintro}) by replacing $g_{\mu \nu}$ with the generalized metric of DFT $\h{\cal H}_{\h M \h N}$. However, the first term is ambiguous in terms of DFT projections, namely
\bea
b_{1} U_{\underline{\h M}} U_{\underline {\h N}} + b_{2} U_{\underline {\h M}} U_{\overline {\h N}} + b_{3} U_{\overline {\h M}} U_{\underline {\h N}} + b_{4} U_{\overline {\h M}} U_{\overline {\h N}} \, .
\label{projectintro}
\eea
The simplest proposal is to consider only mixed components in (\ref{projectintro}) since this directly matches the structure of the generalized energy--momentum tensor coming from the variation of the DFT for the generalized scalar field \cite{phaseDFT}. We fix the remaining coefficients such that
\bea
\h{\cal T}^{\h M \h N} = 2(\tilde e+\tilde p) \h U^{\un {\h M}} \h U^{\ov {\h N}} + 2(\tilde e+\tilde p) \h U^{\ov M} \h U^{\un {\h N}} + \tilde p \h {\cal H}^{\h M \h N} \, ,
\label{emintro}
\eea
where $\tilde e(\h X)=\tilde e$ and $\tilde p(\h X)=\tilde p$ are defined as the generalized energy density and pressure, and $\h U^{\h M}$ is the generalized velocity which can be parametrized as
\bea
\h U^{\h M}= (\tilde{u}_{\mu}, u^{\mu}) \, ,
\eea
with $\tilde u_{\mu}$ a dual velocity.

In addition to the construction of the generalization of the energy--momentum tensor for the perfect fluid (\ref{emintro}) in the double geometry, we also explore the conservation equations derived from (\ref{conserv}) and we elaborate on the DFT Einstein--type equations (\ref{einsteindftintro}) sourced by (\ref{emintro}) for a cosmological ansatz and their parametrization. As we have explained, this proposal must agree with the previous cosmological DFT constructions such as \cite{ParkCosmo} and \cite{Wu}, upon parametrization. Considering the space--time split and a rest frame the non--trivial equations coming from the generalized Einstein equation are given by 
\bea
\label{introR}
\boxed{\begin{split}
	\label{CosmoDFTintro}
	-\tilde p & =  2 (\partial_{0}{d})^2 - 2 \partial_{00}{d} - \frac{1}{16} \partial_{0}{A^{M N}} \partial_{0}{A_{M N}}    + (\partial_0 \leftrightarrow \tilde\partial^0)\\
	\frac12(\tilde e+\tilde p) & =      \partial_{00}d + \frac{1}{16} \partial_0 A^{KL} \partial_0 A_{KL} - (\partial_0 \leftrightarrow \tilde\partial^0)\\
	0 & = - P_{M}{}^{P} \left[\frac14 \partial_{00} A_{PQ}  - \frac12 \partial_0 d \partial_0 A_{PQ} + (\partial_0 \leftrightarrow \tilde\partial^0) \right] \ov{P}^{Q}{}_{N} + \left(P \leftrightarrow \ov P \right).\quad
\end{split}}
\eea
These equations determine the set of DFT cosmologies whose matter source is (\ref{emintro}) and their parametrization reads 
\bea
\label{Cosmosugraintro}
\boxed{\begin{split}
	-2\tilde p & =  2 (D-1) \dot{H} + D(D-1) H^{2} - 4 \ddot{\phi} + 4 \dot{\phi}^2 - 4 (D-1) H \dot{\phi}  \\ & + \frac1{4\,a^{4}} \delta^{i j} \delta^{k l} \dot b_{i k} \dot b_{j l}\\ 
	-(\tilde e+\tilde p)& =  (D-1) \left(\dot{H} + H^{2}\right) -2\ddot{\phi} + \frac1{4\,a^4} \delta^{i j} \delta^{k l} \dot b_{i k} \dot b_{j l} \\
	0  & =   \left(\dot H + (D-1) H^{2}  - 2 H \dot{\phi}  \right) \delta^{i j} + \frac1{2\,a^4} \delta^{i k} \delta^{j l} \delta^{m n} \dot b_{k m} \dot b_{l n} 
	\\
	0 & =   \delta^{j l} \ddot b_{i l} + \frac{1}{a^4} \delta^{j k} \delta^{l m} \delta^{n o} \dot b_{l n} \dot b_{k m} b_{i o}  +\left[(D-5)H -2\dot\phi\right]  \delta^{j k} \dot b_{i k}\ \\
	& - \left[2 \dot{H} + 2(D-1)H^{2} - 4 H  \dot \phi \right] \delta^{j k}  b_{i k} \, .
\end{split}}
\eea
Here we keep the usual coordinates and eliminate the dual ones. However, other T--dual solutions can be easily obtained from (\ref{introR}). These cosmologies reduce to the results of \cite{Wu} when the matter and the $b$--field are neglected. Furthermore it is possible to describe string cosmologies scenarios \cite{Gasperini} by implementing the following field redefinitions 
\bea
\tilde p = e^{2 \phi} p \, , \quad
\tilde e =  e^{2\phi} e 
\label{redefintro}\, .
\eea

The succeed of DFT with matter yields in the fact that it is possible to obtain string cosmologies from the double geometry, whose only stringy ingredient is given by the duality group invariance. As we discuss in Section \ref{fromdft}, it is worth noting that the proposal (\ref{emintro}) is a first step in the description of the generalized perfect fluid in the double geometry, which reproduces string cosmologies with a fixed dilaton source $\sigma$ (cf. (\ref{sigma})) upon a suitable parametrization and imposing the strong constraint.

\subsection{Outline}

This paper is organized as follows: In Section \ref{DFT} we introduce the basics of DFT and we describe the space--time split decomposition. In Section \ref{Incorporating} we incorporate matter into DFT: firstly we describe the construction of the DFT phase space from a kinetic theory point of view. Secondly we propose an explicit form for the generalized energy--momentum tensor related to a perfect fluid in the double geometry. Finally we study the conservation laws from the divergence equation. We analyze the cosmological ansatz of DFT with matter given by the generalized Einstein equation, in Section \ref{Solutions}. We also show the agreement with string cosmology considering field redefinitions of the DFT energy density and pressure. Conclusions are given in Section \ref{Conclusions}.

\section{Vacuum double field theory}
\label{DFT}
In this section we introduce the basic aspects of the vacuum sector of DFT. We start by considering a formulation where the fundamental fields are the generalized metric and the generalized dilaton, which are in $O(D,D)$ representations. Then, we describe the space--time split framework. In both approaches, the fundamental fields depend on both the ordinary coordinates and the dual ones.

\subsection{Basics} 
DFT \cite{Siegel,DFT,DFTkorea} is a proposal to rewrite the low energy limit of ST as a manifestly $O(D,D)$ invariant theory. All the DFT fields and parameters are duality invariant objects. The double geometry consists in a $2D$--dimensional space with coordinates $\h X^{\hat M}=(\tilde{x}_{\mu},x^{\mu})$, $\hat M=1,\dots,2D$. Here $x^{\mu}$ are coordinates on an embedded supergravity, and $\tilde{x}_{\mu}$ are dual coordinates. Considering a solution to the weak and strong constraints, namely
\bea
\partial_{\hat M} (\partial^{\hat M} \star) & = & 0 \nn \\ (\partial_{\hat M} \star) (\partial^{\hat M} \star) & = & 0 \, 
\label{SC}
\eea
where $\star$ is a generic field/parameter, half of the coordinates are taken away. For instance, simple solutions are $\tilde \partial^{\mu}=0$ or $\partial_{\mu}=0$ in which the fundamental fields depend only on $x^{\mu}$ or $\tilde{x}_{\mu}$, respectively. In (\ref{SC}), contractions are given by the $O(D,D)$ invariant metric, $\hat \eta_{\hat M \hat N}$. 

The theory is invariant under a global $O(D,D)$ symmetry which infinitesimally reads
\bea
\delta V_{\h M} = V_{\h N} \,\Omega^{\h N}{}_{\h M} \, ,
\label{duality}
\eea
where $V_{\h M}$ is a generic $O(D,D)$--multiplet and $\Omega \in O(D,D)$ is the generic parameter of the transformation. A generalized notion of diffeomorphisms can be defined in the double space. These are given by the generalized Lie derivative,
\bea
\delta_{\hat \xi} V^{\hat M} = {\mathcal L}_{\hat \xi} V^{\hat M} =\hat \xi^{\hat N} \partial_{\hat N} V^{\hat M} + (\partial^{\hat M} \hat \xi_{\hat P} - \partial_{\hat P} \hat \xi^{\hat M}) V^{\hat P} + \omega \partial_{\hat N} \hat \xi^{\hat N} V^{\hat M}\, ,
\eea
where $\hat \xi^{\hat M}$ a generic parameter and $\omega$ a density weight factor. The closure of these transformations is given by the C--bracket
\bea
\Big[\delta_{\hat \xi_1},\delta_{\hat \xi_2} \Big] V^{\hat M} = \delta_{\hat \xi_{21}} V^{\hat M} 
\eea
where
\bea
\hat \xi^{\hat M}_{12} = \hat \xi^{\hat P}_{1} \frac{\partial \hat \xi^{\hat M}_{2}}{\partial \h X^{\hat P}} - \frac12 \hat \xi^{\hat P}_{1} \frac{\partial \hat \xi_{2\hat P}}{\partial \h X_{\hat M}} - (1 \leftrightarrow 2) \, .
\label{Cbra}
\eea

The fundamental fields are the generalized dilaton  $\h d$ and the generalized metric $\hat{\cal H}_{\hat M \hat N}$. In this work we consider that these fields integrate the vacuum sector of DFT. While the generalized metric transforms as a tensor under $O(D,D)$ transformations and generalized diffeomorphisms ($\omega=0$), and it is an element of the duality group, i.e.
\bea
\hat{\cal H}_{\h M \h P} \hat \eta^{\h P \h Q} \h{\cal H}_{\h Q \h N} = \h \eta_{\h M \h N} \, , 
\eea
the generalized dilaton is an $O(D,D)$ scalar and transforms non--covariantly under generalized diffeomorphisms
\bea
\delta_{\hat \xi} \hat d & = & \hat \xi^{\hat N} \partial_{\hat N} \hat d - \frac12 \partial_{\hat M} \hat \xi^{\hat M} \, .
\eea
Both $\h {\cal H}_{M N}$ and $\h \eta_{M N}$ can be used to construct DFT projectors
\bea
P_{\h M \h N} = \frac{1}{2}\left(\h \eta_{\h M \h N} - \h{\cal H}_{\h M \h N}\right)  \ \ {\rm and} \ \
\ov{P}_{\h M \h N} = \frac{1}{2}\left(\h \eta_{\h M \h N} + \h{\cal H}_{\h M \h N}\right)\ .
\eea

The action principle of DFT is given by 
\bea
S_{DFT} = \frac12 \int d^{2D}\h X e^{-2 \h d} {\cal L}_{DFT} \,\label{actionvacuumdft}
\eea
with the duality invariant Lagrangian
\bea
{\cal L}_{DFT} & = &  \frac18 \h{\cal H}^{\h M \h N} \partial_{\h M}{\h{\cal H}^{\h K \h L}} \partial_{\h N}{\h{\cal H}_{\h K \h L}} - \frac12 \h{\cal H}^{\h M \h N} \partial_{\h N}{\h{\cal H}^{\h K \h L}} \partial_{\h L}{\h {\cal H}_{\h M \h K}} \nn \\
&& + 4 \h{\cal H}^{\h M \h N} \partial_{\h M}{d} \partial_{\h N}{\h d} - 2 \partial_{\h M}{\h {\cal H}^{\h M \h N}} \partial_{\h N}{\hat d} \, .
\eea
Since this Lagrangian transforms as a scalar under generalized diffeomorphisms, the action principle of DFT is invariant.

In DFT it is also possible to construct a generalized version of the Ricci scalar, i.e.,
\bea
\hat {\cal R} = && \frac18 \hat{\cal H}^{\hat M \hat N} \partial_{\hat M}\hat{\cal H}^{\hat K \hat L}\partial_{\hat N}\hat{\cal H}_{\hat K \hat L} - \frac12 \hat{\cal H}^{\hat M \hat N}\partial_{\hat N}\hat{\cal H}^{\hat K \hat L}\partial_{\hat L}\hat{\cal H}_{\hat M \hat K} + 4 \hat{\cal H}^{\hat M \hat N} \partial_{\hat M}\partial_{\hat N} \hat d \nn \\ && + 4 \partial_{\hat M}\hat{\cal H}^{\hat M \hat N} \partial_{\hat N}\hat d - 4 \hat{\cal H}^{\hat M \hat N} \partial_{\hat M}\hat d \partial_{\hat N}\hat d -  \partial_{\hat M} \partial_{\hat N} \hat{\cal H}^{\hat M \hat N} \, , \label{ricciscalar}
\eea
and a generalized Ricci tensor
\bea
\h {\cal R}_{\h M \h N} = P_{\h M}{}^{\h P} \h{\cal K}_{\h P \h Q} {\bar P}^{\h Q}{}_{\h N} +  {\bar P}_{\h M}{}^{\h P} \h{\cal K}_{\h P \h Q} P^{\h Q}{}_{\h N} \, ,\label{riccitensor}
\eea
where
\bea
\h {\cal K}_{\h M \h N} & = & \frac{1}{8} \partial_{\h M} \h{\cal H}^{\h K \h L} \partial_{\h N} \h{\cal H}_{\h K \h L} - \frac14 \left(\partial_{\h L} - 2 \partial_{\h L} \h d\right)\left(\h{\cal H}^{\h L \h K} \partial_{\h K}\h{\cal H}_{\h M \h N}\right) + 2 \partial_{\h M}\partial_{\h N} \h d \nn \\ && - \frac12 \partial_{(\h M} \h{\cal H}^{\h K \h L} \partial_{\h L} \h{\cal H}_{\h N) \h K} + \frac12 \left(\partial_{\h L} - 2 \partial_{\h L} \h d\right) \left(\h {\cal H}^{\h K \h L} \partial_{(\h M} \h{\cal H}_{\h N) \h K} + \h{\cal H}^{\h K}{}_{(\h M} \partial_{\h K} {\cal H}^{\h L}{}_{\h N)}\right) \, . \nn 
\eea
It is straightforward to verify that these objects are fully covariant. Moreover, they can be constructed from a generalized Riemann tensor, but the latter cannot be entirely expressed in terms of the fundamental fields of DFT \cite{RiemannDFT}.

\subsection{Space--time split of double field theory}\label{split}
We start by splitting the DFT coordinates as
\bea
\hat{X}^{\hat M} = (\tilde x_{0}, x^{0}, X^{M})\label{coordinatesplit}
\eea
where $ M=3,\dots,2D$, $\tilde x_{0}=\tilde t$, $x^{0}=t$ and the partial derivatives are \bea
\partial_{\hat M} = (\tilde \partial^{0}, \partial_{0}, \partial_M) \, . 
\eea
The weak constraint now implies
\bea
\partial_{M}\partial^{M}\star = - 2 \tilde \partial^{0}\partial_{0}\star
\label{weak}
\eea
while the strong constraint is
\bea
\partial_{M} \Box \partial^{M} \star = - \partial_{0} \Box \tilde\partial^{0} \star - \partial_{0} \star \tilde\partial^{0} \Box \, ,
\eea
where $\star$ and $\Box$ are arbitrary fields/parameters. The $O(D,D)$ invariant metric decomposes as
\bea
\hat{\eta}_{\hat M \hat N}  = \left(\begin{matrix}\hat{\eta}^{00}&\hat{\eta}^{0}{}_{0} & \hat{\eta}^{0}{}_{N}\\ 
\hat{\eta}_{0}{}^{0}& \hat{\eta}_{0 0} & \hat{\eta}_{0 N} \\ \hat{\eta}_{M}{}^{0} & \hat{\eta}_{M 0} & \hat{\eta}_{MN} \end{matrix}\right) = \left(\begin{matrix}0&1 & 0\\ 
1&0 & 0 \\ 0 & 0 & \eta_{MN} \end{matrix}\right) \, , 
\label{etasp}
\eea 
while the generalized metric is given by
\footnotesize
\bea
\hat{\cal H}_{\hat M \hat N} & = &  \left(\begin{matrix}-N^{-2} & \alpha & N^{-2} {\cal N}_{N}\\ 
\alpha & -\frac12 \alpha {\cal N}^{K} {\cal N}_{K}-N^{2} + {\cal H}_{PK} {\cal N}^P {\cal N}^K & - \alpha {\cal N}_{N} + {\cal H}_{N K} {\cal N}^K \\ N^{-2} {\cal N}_{M} & - \alpha {\cal N}_{M} + {\cal H}_{M K} {\cal N}^{K} & {\cal H}_{MN} - N^{-2} {\cal N}_{M} {\cal N}_{N} \end{matrix}\right) \, 
\label{Hsp}
\eea 
\normalsize
where $\alpha=\frac12 N^{-2} {\cal N}^M {\cal N}_{M}$ \cite{Naseer}. In (\ref{Hsp}) ${\cal N}_{M}$ is a generalized shift vector and $N$ is a generalized lapse function. The generalized dilaton, $\hat{d}$, can be redefined in terms of the generalized lapse function,
\bea
e^{-2 \hat d} = N e^{-2 d} \, .
\eea
Since we are interested in cosmological ansatz, we consider ${\cal N}_{M}=0$ and $N=1$. So far, the line element is given by
\bea
dS^2 = - d\tilde t^2 - dt^2 + {\cal H}_{MN} dX^{M} dX^{N} \, , 
\eea
with ${\cal H}_{MN}={\cal H}_{M N}(\tilde t, t, X)$. The dependence in all the dual coordinates is crucial to preserve the $O(D,D)$ invariance. Since we are considering $N=1$, the generalized dilaton is not redefined and we get $\h d(\h X) = d(\tilde t, t, X)$.

We can easily decompose both the DFT Lagrangian \bea
{\cal L}_{DFT} & = & \frac18 {\cal H}^{P Q} \partial_{P}{{\cal H}^{M N}} \partial_{Q}{{\cal H}_{M N}} - \frac18 \partial_{0}{{\cal H}^{M N}} \partial_{0}{{\cal H}_{M N}} - \frac18 \tilde \partial^{0}{{\cal H}^{M N}} \tilde \partial^{0}{{\cal H}_{M N}} \nn \\ && - \frac12 {\cal H}^{P Q} \partial_{Q}{{\cal H}^{M N}} \partial_{N}{{\cal H}_{P M}}   + 4 {\cal H}^{M N} \partial_{M}{d} \partial_{N}{d} - 4 \partial_{0}{d} \partial_{0}{d} \nn \\ && - 4 \tilde \partial^{0}{d} \tilde \partial^{0}{d} - 2 \partial_{M}{{\cal H}^{M N}} \partial_{N}{d} \, ,
\eea
and the generalized curvatures
\bea
\h {\cal R} & = & \frac18 {\cal H}^{P Q} \partial_{P}{{\cal H}^{M N}} \partial_{Q}{{\cal H}_{M N}} - \frac18 \partial_{0}{{\cal H}^{M N}} \partial_{0}{{\cal H}_{M N}} - \frac18 \tilde \partial^{0}{{\cal H}^{M N}} \tilde \partial^{0}{{\cal H}_{M N}} \nn \\ && - \frac12 {\cal H}^{P Q} \partial_{Q}{{\cal H}^{M N}} \partial_{N}{{\cal H}_{P M}} + 4 {\cal H}^{M N} \partial_{M N}{d} - 4 \partial_{00}{d} \nn \\ && - 4 \tilde \partial^{00}{d} + 4 \partial_{M}{{\cal H}^{M N}} \partial_{N}{d} - 4 {\cal H}^{M N} \partial_{M}{d} \partial_{N}{d} \nn \\ && + 4 \partial_{0}{d} \partial_{0}{d} + 4 \tilde \partial^{0}{d} \tilde \partial^{0}{d} - \partial_{M N}{{\cal H}^{M N}} \, 
\eea
and
\bea
\h{\cal K}_{\hat M \hat N} & = & \frac18 \partial_{\h M} {\cal{\h H}}^{K L} \partial_{\hat N} {\cal{\h H}}_{KL} - \frac14 \partial_{L} {\cal{\h H}}^{K L} \partial_{K}\h{\cal H}_{\hat M \hat N} + \frac14 \partial_{00} \h{\cal H}_{\h M \h N} + \frac14 \tilde \partial^{00} \h{\cal H}_{\h M \h N} \nn \\ && - \frac14 {\cal{\h H}}^{LK} \partial_{LK} \h{\cal H}_{\h M \h N}  -\frac12 \partial_{0} d \partial_{0} \h{\cal H}_{\h M \h N} - \frac12 \tilde \partial^0 d \partial^0 \h{\cal H}_{\h M \h N} + \frac12 \partial_L d {\cal H}^{LK} \partial_{K}\h{\cal H}_{\h M \h N} \nn \\ && + 2 \partial_{\h M \h N} d  - \frac12 \partial_{(\hat M} {\cal H}^{KL} \partial_{L} \h{\cal H}_{\h N) K} + \frac12 \partial_{L} {\cal H}^{KL} \partial_{(\h M} \h{\cal H}_{\h N)K} + \frac12 {\cal H}^{KL} \partial_{L(\h M} \h{\cal H}_{\h N) K} \nn \\ && - \partial_{L}d {\cal H}^{KL} \partial_{(\h M} \h{\cal H}_{\h N)K}  + \frac12 \partial_{L} \h{\cal H}^{K}{}_{(\h M} \partial_{K} \h{\cal H}^{L}{}_{\h N)} + \frac12 \h{\cal H}_{0 (\h M} \partial_{L} \tilde \partial^{0} \h{\cal H}^{L}{}_{\h N)} \nn \\ && + \frac12 \h{\cal H}^{0}{}_{(\h M} \partial_{L} \partial_{0} \h{\cal H}^{L}{}_{\h N)} + \frac12 \h{\cal H}^{K}{}_{(\h M} \partial_{LK} \h{\cal H}^{L}{}_{\h N)} - \frac12 \partial_L d \h{\cal H}^{0}{}_{(\h M} \partial_{0} \h{\cal H}^{L}{}_{\h N)} \nn \\ && - \frac12 \partial_L d \h{\cal H}_{0(\h M} \tilde \partial^{0} \h{\cal H}^{L}{}_{\h N)} - \frac12 \partial_L d \h{\cal H}^{K}{}_{(\h M} \partial_{K} \h{\cal H}^{L}{}_{\h N)} \, .
\eea
The previous expressions are useful to obtain closed expressions for rewriting cosmological solutions in a duality invariant way in the DFT with matter framework. Moreover, considering a cosmological principle \cite{Wu}, we can impose ${\cal H}_{MN}(\tilde t,t, X)=A_{MN}(\tilde t,t)$ and $d(\tilde t,t, X)=d(\tilde t,t)$ and the expressions for the curvatures drastically simplify. In this case, the remaining symmetry is $O(1,1) \times O(D-1,D-1)$. We discuss about this point in Section \ref{Solutions}.

\section{Incorporating matter into double field theory}
\label{Incorporating}

The original construction of DFT \cite{DFT} is based on a rewriting of the low energy limit of ST where the invariance under $O(D,D)$ is accomplished before compactification. In fact, the dynamics of the supergravity NS-NS vacuum sector regarding the metric $g_{\mu\nu}$, the $b$--field and the dilaton $\phi$ can be completely rewritten within the framework of DFT and, particularly, in terms of $O(D,D)$ multiplets. In this case a generalized vacuum Einstein equation can be constructed as
\bea
\h {\cal G}_{\hat M \hat N} = 0 \, ,\label{einsteinvacuum}
\eea
where the LHS is given by the symmetric generalized Einstein tensor 
\bea
\hat {\cal G}_{\hat M \hat N} = - \frac12 \hat {\cal H}_{\hat M \hat N} \hat {\cal R} - 2 \h {\cal R}_{\h M \h N} \, ,
\eea
with $\hat {\cal R}$ and $\h {\cal R}_{\h M \h N}$ given in equations (\ref{ricciscalar}) and (\ref{riccitensor}) respectively. In order to get the full dynamics, equation (\ref{einsteinvacuum}) must be complemented with the equation of motion coming from the variation of the vacuum action (\ref{actionvacuumdft}) with respect to the dilaton, namely $\hat {\cal R}=0$.

The inclusion of matter in the dynamics at the level of supergravity is also well understood both through a variational principle from an specific matter action or through the usual Einstein equation with a proper energy--momentum tensor. While the relation between supergravity, matter and T--duality was studied in several works, the generalized Einstein equation in terms of $O(D,D)$ multiplets was constructed in \cite{ParkCosmo} and takes the following form
\bea
\h {\cal G}_{\hat M \hat N} = \h {\cal T}_{\hat M \hat N} \, ,
\label{EinsDFT}
\eea
where $\h {\cal T}_{\hat M \hat N}$ is the generalized symmetric energy--momentum tensor which incorporates the effects of matter into the dynamics at the level of DFT.

The aim of this section is to find an explicit expression for $\h {\cal T}_{\hat M \hat N}$ written only in terms of manifestly $O(D,D)$--covariant quantities that recovers the well-known string cosmology upon suitable parametrization. So that we propose the inclusion of matter into the DFT formulation based on kinetic theory extending the results of \cite{phaseDFT}, where the generalized energy--momentum tensor was defined as the second moment of a generalized one--particle distribution function. 

In the first part of this section we give a brief review of the phase space for the point particle in the double geometry. Then we construct the generalized energy--momentum tensor for a perfect fluid in the double geometry. This proposal is given in terms of a generalized velocity, which is consistently constrained. We present the conservation laws for the generalized energy--momentum tensor which are given by the generalized Euler equations and the generalized relativistic energy conservation equation.

\subsection{Double phase space and the point particle}

The phase space of DFT can be constructed considering an extension of the double geometry with coordinates, 
\bea
\Big\{\h X^{\h M},\h {\cal P}^{\h M} \Big\} \, ,
\eea
where $\h {\cal P}^{\h M}$ is the generalized momentum. These coordinates are defined on a double tangent space, they transform as vectors with respect to $O(D,D)$ and satisfy 
\bea
\frac{\partial \h {\cal P}^{\h M}}{\partial \h X^{\h N}} = 0 \, .
\eea
The strong constraint must also be extended as
\bea
\label{SC1}
\left(\frac{\partial}{\partial \h {\cal P}^{\h M}} \star\right) \left(\frac{\partial}{\partial \h {\cal P}_{\h M}} \star\right) = \frac{\partial}{\partial \h {\cal P}^{\h M}} \left(\frac{\partial}{\partial \h {\cal P}_{\h M}} \star\right) & = & 0 \, , \\
\left(\frac{\partial}{\partial \h X^{\h M}} \star\right) \left(\frac{\partial}{\partial \h {\cal P}_{\h M}} \star\right) = \frac{\partial}{\partial \h X^{\h M}} \left(\frac{\partial}{\partial \h {\cal P}_{\h M}} \star\right) & = & 0 \, .   
\label{SC2}
\eea
Equations (\ref{SC1}) and (\ref{SC2}) guarantee that the generalized diffeomorphisms on the phase space, 
\bea
\delta_{\h \xi} V^{\h Q}(\h X,\h {\cal P}) = {\cal L}_{\h \xi} V^{\h Q}(\h X,\h {\cal P}) + \h {\cal P}^{\h N} \frac{\partial \h \xi^{\h M}}{\partial \h X^{\h N}} \frac{\partial \h V^{\h Q}(\h X,\h {\cal P})}{\partial \h {\cal P}^{\h M}} - \h {\cal P}^{\h N} \frac{\partial \h \xi_{\h N}}{\partial \h X_{\h M}} \frac{\partial V^{\h Q}(\h X,\h {\cal P})}{\partial \h {\cal P}^{\h M}}  \, ,
\label{trans}
\eea
satisfy a closure condition, 
\bea
\Big[\delta_{\h \xi_1},\delta_{\h \xi_2} \Big] V^{\h M}(\h X,\h {\cal P}) = -\delta_{\h \xi_{12}} V^{\h M}(\h X,\h {\cal P}) \,,
\eea
where $\h \xi^{M}_{12}(\h X)$ is given by the C--bracket (eq. (\ref{Cbra})). In (\ref{trans}) each diffeomorphism parameter depends only on the space--time coordinates, $\h \xi^{\h M}=\h \xi^{\h M}(\h X)$, and $V^{\h Q}(\h X,\h {\cal P})$ is a generic vector on the double phase space.

The generalized energy--momentum tensor is a tensor in the double space,
\bea
\hat{\cal T}^{\hat M \hat N}(\h X) = \int \h{\cal P}^{\hat M} \h{\cal P}^{\h N} \h F\, e^{-2\h d} d^{2D}\h{\cal P} \, ,\label{emt}
\eea
where $\h F=\h F(\h X,\h {\cal P})$ is the one--particle generalized distribution function. This function transforms as a phase space scalar. Indeed the generalized energy--momentum tensor fulfills a conservation--like equation
\bea
\nabla_{\h M}\h {\cal T}^{\h M \h N} = 0  \, ,
\label{clawDFT}
\eea
as it is shown in \cite{phaseDFT} by computing the divergence of the second moment of the generalized Boltzmann equation for an equilibrium state. 

On the other hand, the generalized momentum of a particle encodes the physics of an ordinary D--dimensional momentum $p_{\mu}=g_{\mu \nu} p^{\nu}$ so we need to impose a strong--constraint like equation, 
\bea
\h{\cal P}^{\h M} \h{\eta}_{\h M \h N} \h{\cal P}^{\h N} & = & 0 \, ,
\label{constraints1}
\eea
in order to get the correct number of degrees of freedom when parametrizing. In fact the equation (\ref{constraints1}) takes the same form as the level matching condition in toroidal compactifications \cite{ReviewDFT1}. Similarly, the 
mass--shell like condition is given by
\bea
-\h{\cal P}^{\h M} \h{\cal H}_{\h M \h N} \h{\cal P}^{\h N} =  m^2 \, .
\label{constraints2}
\eea
which takes the form of the mass--squared operator. When ${\cal P}^{\h M}$ arranges the $D$--dimensional momentum and winding modes for a closed string in a toroidal background, this object is an $O(D,D,Z)$ multiplet \cite{Porrati} and the equation (\ref{constraints2}) reproduces the spectrum of the massless states of the string. Here the generalized momentum describes the momentum of a particle in the double geometry and, therefore, it is an $O(D,D,R)$ multiplet. Furthermore, the parametrization of the phase-space diffeomorphism (\ref{trans}) requires $\h{\cal P}^{\h M}=(\tilde p^{\mu},p_{\mu})=(0,p_{\mu})$ which is a solution of (\ref{constraints1}) and in turn gives rise to the ordinary mass--shell condition,
\bea
-p_{\mu} g^{\mu \nu} p_{\nu} = m^2 \, ,
\eea
through equation (\ref{constraints2}).

Additionally the generalized velocity for the particle in the double geometry is given by
\bea
\hat U^{\hat M} = \frac{D\hat X^{\hat M}}{D\tau} \, ,\label{u}
\eea
where $DX^M$ is DFT compatible 1--form differential \cite{Prk} and $\tau$ is an affine parameter in the double geometry, which reduces to the standard proper time upon parametrization. Both the generalized momentum $\hat {\mathcal P}^{\hat M}$, evaluated on the trajectory of a point--particle, and the generalized velocity $\hat U^{\hat M}$ are vectors that indicate the same geometric direction in the tangent space and then they must be proportional. Choosing the proportionality constant is, in fact, defining an specific affine parameter $\tau$ in (\ref{u}). Indeed, we take the relation 
\bea
\hat {\mathcal P}^{\hat M}=m\,\hat U^{\hat M}=m\,D\hat X^{\hat M}/D\tau \, , 
\eea
where $m$ is the invariant mass in ($\ref{constraints2}$).

It is straightforward to rewrite the properties of the generalized momentum in terms of the generalized velocity as
\begin{eqnarray}
\label{Uprop1}
\hat U_{\hat M}  \h{\cal H}^{\hat M\hat N}\hat U_{\hat N}&=&-1\\
\hat U_{\hat M}  {\hat \eta}^{\hat M\hat N} \hat U_{\hat N}&=&0 \, .
\label{Uprop2}
\end{eqnarray}

In the next part of this section we take advantage of the phase--space construction of DFT and we give a proposal for the generalized energy--momentum tensor of a perfect fluid in the double space. This is a simple proposal to effectively describe matter contributions in DFT such as fermionic \cite{Susy} or Ramond--Ramond contributions \cite{RR}, among others.

\subsection{The perfect fluid in the double geometry}

A relevant ingredient for including a perfect fluid in DFT is the generalized symmetric energy--momentum tensor $\mathcal{T}_{\hat M\hat N}$. In our case its index structure is given only by the fundamental 2--index tensors of DFT, $\hat{\mathcal{H}}_{\hat M\hat N}$ and $\hat \eta_{\hat M\hat N}$, and the generalized velocity vector of the fluid $\hat U_{\hat M}$ which defines the temporal direction upon the split (\ref{coordinatesplit}). Here we give a first proposal inspired by the structure of the ordinary energy--momentum tensor in the Riemannian geometries, i.e., 
\bea
T_{\mu \nu}=(p+e) u_{\mu} u_{\nu} + p g_{\mu \nu} \, .\label{tmunurelativista}
\eea
While the second term of the previous expression can be easily generalized replacing $g_{\mu \nu}$ by $\h{\cal H}_{\h M \h N}$, the first term is ambiguous in terms of the DFT projections,
\bea
b_{1} U_{\underline{\h M}} U_{\underline {\h N}} + b_{2} U_{\underline {\h M}} U_{\overline {\h N}} + b_{3} U_{\overline {\h M}} U_{\underline {\h N}} + b_{4} U_{\overline {\h M}} U_{\overline {\h N}} \, .
\label{project}
\eea

The dynamics of the perfect fluid in Riemannian geometries is equivalent to the dynamics of a scalar field \cite{scalarfluidgr}. Since the dynamics for the generalized scalar field is given by a straightforward generalization of the Klein--Gordon equation it is expected that the minimal proposal for the generalized energy--momentum of the perfect fluid contains non--trivial $b_{2}$ and $b_{3}$, which are present in the generalized energy--momentum tensor of the generalized scalar field (see eq. (4.24) in \cite{phaseDFT}). Therefore our proposal for the generalized energy--momentum tensor reads   
\bea
\h{\cal T}_{\h M \h N}= B\,\big(\h U_{\h{\un M}} {\h U}_{\ov{\hat N}} + \,\h U_{\ov{\h M}} {\h U}_{\h{\un N}}\big) + C\,\mathcal{{\h H}}_{\h M \h N}\, ,
\label{start}
\eea
where we have neglected the $b_{1}$ and $b_{4}$ and possible dilatonic contributions. Each term of (\ref{start}) transforms covariantly under T--duality through (\ref{duality}), so these transformations mix the components of the energy--momentum tensor as pointed out in \cite{GasVene}. The expression (\ref{start}) hence capture all these (T--dual) tensors in a unified approach. 

Using the space--time split formulation of DFT, the generalized velocity $\hat U_{\hat M}$ splits according to
\bea
\h U^{\h M} = (\tilde{u}_{0}, u^{0},U^M) \, .
\eea
Furthermore, from the relations (\ref{Uprop1})--(\ref{Uprop2}) we get
\bea
\h U_{\ov {\h M}} \h U^{\ov{\h M}} = - \frac12 \, , \quad \quad
\h U_{\un{\h M}} \h U^{\un{\h M}}  =  \frac12\,.
\eea
Finally, we identify the quantities $B$ and $C$ in the generalized energy--momentum tensor (\ref{start}) as
\bea
B & = & 2\big[\tilde e(X)+ \tilde p(X)\big] \nn \\
C & = & \tilde p(X) \, .\label{prhotilde}
\eea
Here we define $\tilde e(X)$ and $\tilde p(X)$ as the generalized notions of the energy density and pressure from a formal analogy between (\ref{start}) and the structure of the usual energy--momentum tensor of the relativistic perfect fluid (\ref{tmunurelativista}). The tilded variables in these expressions may be related to the ordinary energy density and pressure through a field redefinition as we show in Section \ref{fromdft}.

On the other hand, we extend the DFT projectors in the following way,
\bea
\ov h_{\h M \h N} & = & \ov P_{\h M \h N} + 2 \ov P_{\h M}{}^{\h P} \ov P_{\h N}{}^{\h Q} \h U_{\h P} \h U_{\h Q} \, ,\\
\un h_{\h M \h N} & = & P_{\h M \h N} - 2 P_{\h M}{}^{\h P} P_{\h N}{}^{\h Q} \h U_{\h P}  \h U_{\h Q} \, ,
\eea
and therefore $\ov h_{\h M \h P} \ov h^{\h P}{}_{\h N}=\ov h_{\h M \h N}$ and $\un h_{\h M \h P} \un h^{\h P}{}_{\h N}=\un h_{\h M \h N}$. These projectors also satisfy the orthogonality conditions
\bea
\ov h_{\h M \h N} \h U^{\h M} & = & 0 \, , \\
\un h_{\h M \h N} \h U^{\h M} & = & 0 \, .
\eea

The conservation law for the generalized energy--momentum tensor was deduced in \cite{phaseDFT} and takes the form,   
\bea
\nabla_{\h M}\left( 2(\tilde e+\tilde p) \h U^{\un {\h M}} \h U^{\ov {\h N}} + 2(\tilde e+\tilde p) \h U^{\ov {\h M}} \h U^{\un {\h N}} + \tilde p \h {\cal H}^{\h M \h N} \right)=0 \, .
\label{claw}
\eea
We can extract the generalization of the energy conservation equation for a perfect fluid and the generalization of the relativistic Euler equation in the double space from (\ref{claw}). Indeed the equation (\ref{claw}) may be projected using the generalized velocities, $\h U_{\ov {\h N}}, \h U_{\un {\h N}}$ and the projectors $\ov h_{\h N \h P}, \un{h}_{\h N \h P}$. Thus we get four equations, one per each type of projection.

First we project (\ref{claw}) using the different projections of the generalized velocity and we get
\bea
\h U_{\ov {\h N}} \nabla_{\h M} \left( 2(\tilde e+\tilde p) \h U^{\un {\h M}} \h U^{\ov {\h N}} + 2(\tilde e+\tilde p) \h U^{\ov {\h M}} \h U^{\un {\h N}} + \tilde p \h {\cal H}^{\h M \h N} \right) & = & 0 \, , \\
\h U_{\un {\h N}} \nabla_{\h M} \left( 2(\tilde e+\tilde p) \h U^{\un {\h M}} \h U^{\ov {\h N}} + 2(\tilde e+\tilde p) \h U^{\ov {\h M}} \h U^{\un {\h N}} + \tilde p \h {\cal H}^{\h M \h N} \right) & = & 0 \, .
\eea
Therefore, the generalized relativistic energy conservation equation reads
\footnotesize
\bea
-(\tilde e+\tilde p) \nabla_{\un{\h M}} \h U^{\un{\h M}} -U^{\h{\un M}} \nabla_{\h{\un M}}(\tilde e+\tilde p) + 2 (\tilde e+\tilde p) \h U_{\ov{\h N}} (\h U^{\ov{\h M}} \nabla_{\ov{\h M}} U^{\un{\h N}} + \h U^{\un{\h M}} \nabla_{\un{\h M}} U^{\ov{\h N}}) & = & - \h U^{\ov {\h M}} \nabla_{\ov {\h M}} \tilde p \, ,  \nn \\ 
-(\tilde e+\tilde p) \nabla_{\ov{\h M}} \h U^{\ov{\h M}} -U^{\h{\ov M}} \nabla_{\h{\ov M}}(\tilde e+\tilde p) - 2 (\tilde e+\tilde p) \h U_{\un{\h N}} (\h U^{\un{\h M}} \nabla_{\un{\h M}} U^{\ov{\h N}} + \h U^{\ov{\h M}} \nabla_{\ov{\h M}} U^{\un{\h N}}) & = & - \h U^{\un {\h M}} \nabla_{\un {\h M}} \tilde p \, . \nn
\label{Energy}
\eea
\normalsize

On the other hand the projection (\ref{claw}) with $\ov h_{\h N \h P}$ is
\bea
\ov h_{\h N \h P} \nabla_{\h M} \left( 2(\tilde e+\tilde p) \h U^{\un {\h M}} \h U^{\ov {\h N}} + 2(\tilde e+\tilde p) \h U^{\ov {\h M}} \h U^{\un {\h N}} + \tilde p \h {\cal H}^{\h M \h N} \right) = 0 \, .
\eea
Consequently, the generalized Euler equation projected onto $\ov h_{\h N \h P}$, written in a duality invariant way, is
\bea
2 (\tilde e+\tilde p) \ov h_{\h N \h P} \Big(\h U^{\ov {\h M}} \nabla_{\ov {\h M}} \h U^{\un {\h N}} + \h U^{\un {\h M}} \nabla_{\un {\h M}} \h U^{\ov {\h N}}  \Big) = - \ov h_{\h N \h P} \nabla^{\ov {\h N}} \tilde p \, .
\label{Euler1}
\eea
Analogously we obtain the complementary generalized Euler equation projected with $\un h_{\h N \h P}$,
\bea
2 (\tilde e+\tilde p) \un h_{\h N \h P}  \Big(\h U^{\ov {\h M}} \nabla_{\ov {\h M}} \h U^{\un {\h N}} + \h U^{\un {\h M}} \nabla_{\un {\h M}} \h U^{\ov {\h N}}  \Big) = \un h_{\h N \h P} \nabla^{\h N} \tilde p \, .
\label{Euler2}
\eea

So far we have described how to incorporate matter into the standard formulation of DFT through the generalized energy--momentum tensor for a perfect fluid. In the next section we use this framework to explore the cosmological ansatz with matter.

\section{Cosmological ansatz}
\label{Solutions}
DFT is a powerful framework to rewrite the low energy limit of ST in a T-duality covariant way and, particularly, it is possible to make contact with the vacuum solutions of string cosmology using a suitable parametrization of the DFT fields and parameters. The pioneer work \cite{Wu} introduced the cosmological equations for the dilaton and the metric tensor that came from the generalized Einstein tensor, while in \cite{ParkCosmo} a generalized energy--momentum tensor for the matter coupled to the double geometry was included. Our main goal in this work is to achieve the manifestly covariant system of equations that couples the vacuum fields of the double geometry to matter using a cosmological ansatz at the DFT level. Indeed, in this section we elaborate on a common cosmological DFT framework considering the matter source as a perfect fluid through (\ref{start}).

We firstly perform a space--time split of all the multiplets and then we derive the full dynamics regarding the matter and vacuum DFT fields (\ref{eqdft}). Secondly we parametrized the vacuum fields as usual and write down the parametrized cosmological equations including $g_{\mu\nu}$, $\phi$ and $b$--field contributions (\ref{Cosmosugra}). In addition we find a parametrization of the DFT matter description, (\ref{redef1}) and (\ref{redef2}), which is compatible with the actual pressure and energy density in the string cosmology limit \cite{Gasperini}.

\subsection{Space--time split DFT with matter}
Considering the space--time split (\ref{Hsp}) with ${\cal N}_{M}=0$, $N=1$ and a cosmological principle for the fundamental fields, namely ${\cal H}_{M N} = {A}_{M N}(\tilde t, t)$ and $d=d(\tilde t, t)$,  it is straightforward to decompose the generalized Ricci scalar
\bea
\h{\cal R} & = &  - \frac18 \partial_{0}{A^{M N}} \partial_{0}{A_{M N}} - \frac18 \tilde \partial^{0}{A^{M N}} \tilde \partial^{0}{A_{M N}} - 4 \partial_{00}{d} \nn \\ && - 4 \tilde \partial^{00}{d} + 4 \partial_{0}{d} \partial_{0}{d} + 4 \tilde \partial^{0}{d} \tilde \partial^{0}{d}  \, ,
\eea
the non--vanishing components of the generalized Ricci tensor
\bea
\h{\cal R}_{00} & = & - \h{\cal R}^{0 0} = -\frac12 \h{\cal K}^{00} + \frac12 \h{\cal K}_{00}  \nn \\
\h{\cal R}_{M N} & = & P_{M}{}^{P} \h{\cal K}_{PQ} \ov{P}^{Q}{}_{N} + \ov P_{M}{}^{P} \h{\cal K}_{PQ} {P}^{Q}{}_{N} \, , 
\eea
and the non--vanishing components of ${\cal K}_{\h M \h N}$
\bea
\h{\cal K}_{00} & = & \frac18 \partial_{0}A^{KL} \partial_{0}A_{KL} + 2 \partial_{00}d \\
\h{\cal K}^{00} & = & \frac18 \tilde \partial^0 A^{KL} \tilde \partial^0 A_{KL} + 2 \tilde \partial^{00}d \\
\h{\cal K}_{0}{}^{0} & = & \h{\cal K}^{0}{}_{0} = \frac18 \partial_{0}A^{KL} \tilde \partial^0 A_{KL} \\
\h{\cal K}_{MN} & = & \frac14 \partial_{00} A_{MN} + \frac14 \tilde \partial^{00} A_{MN} - \frac12 \partial_0 d \partial_0 A_{MN} - \frac12 \tilde \partial^0 d \tilde \partial^0 A_{MN}  \, ,
\eea
with (\ref{weak}) already imposed.

On the other hand, we consider that the matter is given by a perfect fluid in the double space, i.e., 
\bea
\h {\cal T}^{\h M \h N}= 2(\tilde e+\tilde p) \h U^{\un {\h M}} \h U^{\ov {\h N}} + 2(\tilde e+\tilde p) \h U^{\ov {\h M}} \h U^{\un {\h N}} + \tilde p \h {\cal H}^{\h M \h N} \, .\label{emtpffull}
\eea
We choose a rest frame such that $\h U^{\h M} = (0,1,0)$ and, therefore, the non--vanishing components of the generalized energy--momentum tensor are 
\bea
\h{\cal T}^{00} & = & (\tilde e+\tilde p)-\tilde p \\ 
\h{\cal T}_{00} & = & -(\tilde e+\tilde p)-\tilde p \\
\h{\cal T}_{MN} & = & \tilde p\, A_{M N} \, .
\eea

From (\ref{EinsDFT}), we get 
\bea
\boxed{\begin{split}
	\label{CosmoDFT}
	-\tilde p & =  2 (\partial_{0}{d})^2 - 2 \partial_{00}{d} - \frac{1}{16} \partial_{0}{A^{M N}} \partial_{0}{A_{M N}}    + (\partial_0 \leftrightarrow \tilde\partial^0)\\
	\frac12(\tilde e+\tilde p) & =      \partial_{00}d + \frac{1}{16} \partial_0 A^{KL} \partial_0 A_{KL} - (\partial_0 \leftrightarrow \tilde\partial^0)\\
	0 & = - P_{M}{}^{P} \left[\frac14 \partial_{00} A_{PQ}  - \frac12 \partial_0 d \partial_0 A_{PQ} + (\partial_0 \leftrightarrow \tilde\partial^0) \right] \ov{P}^{Q}{}_{N} + \left(P \leftrightarrow \ov P \right).\quad
\end{split}}\label{eqdft}
\eea
These are the space--time split DFT equations which can be used to determine cosmological solutions with matter. Strictly speaking, the expressions (\ref{eqdft}) are combinations of the different components of the generalized Einstein equation.

\subsection{Parametrization}
In this section we parametrize all the fundamental quantities in order to analyze the dynamics of the usual supergravity fields. Indeed the parametrization of the $O(D-1,D-1)$ generalized metric reads
\bea
A_{{M N}}  = \frac{1}{a^2} \left(\begin{matrix}\delta^{ij}&-\delta^{ik}b_{kj}\\ 
b_{i k}\delta^{k j}&a^4 \delta_{ij}-b_{i k} \delta^{k l} b_{lj} \end{matrix}\right)\, , \label{aparam}
\eea
where the spatial indices are $i,j=1,\dots,D-1$. Every component only depends on the ordinary time $t$ due to the solution to the strong constraint $\tilde \partial^0=0$. Naturally, it is possible to inspect the dual solution imposing $\partial_0=0$, or any other T--dual combination. The parametrization of the generalized dilaton is given by
\bea
e^{-2d} = a^{D-1} e^{-2\phi} \, ,\label{dparam}
\eea
and, consequently, $d=\phi - \frac{(D-1)}{2} \ln(a)$. The spatial DFT projectors reads
\bea
P_{MN}=\frac12(\eta_{MN} - {A}_{MN}) \,, \quad \quad \ov P_{MN}=\frac12(\eta_{MN} + {A}_{MN}) \, \label{spatialprojectors}
\eea
while 
\bea
\eta_{{M N}}  =  \left(\begin{matrix}0&\delta^{i}_{j}\\ 
\delta_{i}^{j}&0 \end{matrix}\right)\, . \label{etaparam}
\eea

Applying the expressions (\ref{aparam})--(\ref{etaparam}) to the system (\ref{CosmoDFT}) is a straightforward computation, but breaking the duality group in terms of $Gl(D)$ representations generates large expressions which are suitable to manage with a software for symbolic  algebraic tensor manipulation \cite{Cadabra}. Using this code, the parametrization of the dynamical DFT cosmological equations is given by 
\bea
\label{Cosmosugra}
\boxed{\begin{split}
	-2\tilde p & =  2 (D-1) \dot{H} + D(D-1) H^{2} - 4 \ddot{\phi} + 4 \dot{\phi}^2 - 4 (D-1) H \dot{\phi}  \\ & + \frac1{4\,a^{4}} \delta^{i j} \delta^{k l} \dot b_{i k} \dot b_{j l}\\ 
	-(\tilde e+\tilde p)& =  (D-1) \left(\dot{H} + H^{2}\right) -2\ddot{\phi} + \frac1{4\,a^4} \delta^{i j} \delta^{k l} \dot b_{i k} \dot b_{j l} \\
	0  & =   \left(\dot H + (D-1) H^{2}  - 2 H \dot{\phi}  \right) \delta^{i j} + \frac1{2\,a^4} \delta^{i k} \delta^{j l} \delta^{m n} \dot b_{k m} \dot b_{l n} 
	\\
	0 & =   \delta^{j l} \ddot b_{i l} + \frac{1}{a^4} \delta^{j k} \delta^{l m} \delta^{n o} \dot b_{l n} \dot b_{k m} b_{i o}  +\left[(D-5)H -2\dot\phi\right]  \delta^{j k} \dot b_{i k}\ \\
	& - \left[2 \dot{H} + 2(D-1)H^{2} - 4 H  \dot \phi \right] \delta^{j k}  b_{i k}
\end{split}}
\eea
where $H=\dot a/a$, and we keep the same notation for the pressure and the energy density at this supergravity level. Equations (\ref{Cosmosugra}) describe the family of cosmologies that can be rewritten in a DFT framework using the generalized Einstein equation (\ref{EinsDFT}) with a generalized perfect fluid as matter. When $\tilde e=0$, $\tilde p=0$ and $b_{ij}=0$, the vacuum solutions exactly agree with \cite{Wu}.

\subsection{From DFT to string cosmology}\label{fromdft}

Here we analyze the relation between the DFT cosmology with matter and string cosmologies. One possible string cosmology scenario that incorporates matter is based on the following action
\begin{eqnarray}
S_{\rm SC}=\int d^Dx\sqrt{-|g|}e^{-2\phi}\left[ R+4\left(\nabla \phi\right)^2\right]+S_M\,,\label{ssc}
\end{eqnarray}
whose fundamental gravitational fields are the metric tensor $g_{\mu\nu}$ and the dilaton $\phi$, while $b_{ij}=0$. Moreover $R$ is the usual Ricci scalar and $S_M$ is the action term of matter. The equations of motion for the gravitational sector come from the variation of the action with respect to the metric tensor $\delta_g S_{\rm SC}=0$ and to the dilaton $\delta_\phi S_{\rm SC}=0$. In both variations the matter term acts as a source, indeed the source $\delta_g S_M$ is related to the energy--momentum tensor $T_{\mu\nu}$ and, analogously, a dilaton charge (or a dilaton source) $\sigma$ is given by
\bea
\sigma = \frac{-1}{\sqrt{-g}}\frac{\delta S_{M}}{\delta\varphi} =- e^{-2\varphi} \left[\frac{\delta L_{\textrm{mat}}}{\delta \varphi} - 2 L_{\textrm{mat}}\right] \, .\label{sigma}
\eea

Regarding the usual cosmological ansatz, the equations of motion in this string cosmology framework become \cite{Gasperini}
\bea
\label{parametrization}
-e^{2\phi}\sigma & =&  2 (D-1) \dot{H} + D(D-1) H^{2} - 4 \ddot{\phi} + 4 \dot{\phi}^2 - 4 (D-1) H \dot{\phi} \\ 
-e^{2\phi}\left(e+\frac{\sigma}{2}\right)& =&  (D-1) \left(\dot{H} + H^{2}\right) -2\ddot{\phi} \\
e^{2\phi}\left( p-\frac{\sigma}{2}\right)  & = &  \dot H + (D-1) H^{2}  - 2 H \dot{\phi} \, .
\eea

It turns out that these equations are compatible with DFT  when 
\bea
\tilde p & = & e^{2 \phi} p \label{redef1} \, , \\
\tilde e & = & e^{2\phi}e 
\label{redef2}
\eea
and $b_{ij}=0$. In this case the consistency between both approaches forces $p=\sigma/2$. This means that there exists a subfamily of string cosmologies that can be written in a duality covariant way. On the other hand, the inclusion of the remaining cosmologies into a DFT framework may require a Lagrangian--fluid correspondence, as in \cite{scalarfluidgr}, since in that case both $\sigma$ (eq. (\ref{sigma})) and the generalized energy--momentum tensor could be explicitly computed from first principles. 

\section{Conclusions and future directions}
\label{Conclusions}

In this work we have extended the standard vacuum DFT construction in order to include matter through a generalized energy--momentum tensor, which mimics the dynamics of a perfect fluid in the double geometry. We propose an ansatz for this tensor (\ref{start}) in terms of a generalized notion of pressure, energy density and velocity, the latter satisfying a strong constraint--like equation and a mass--shell--like condition (\ref{Uprop1})--(\ref{Uprop2}). We explicitly derive the conservation laws for this tensor (\ref{Energy}) and (\ref{Euler1})-(\ref{Euler2}), which enable the possibility to explore aspects of hydrodynamics at the DFT level.

We perform a space--time split where the dependence of the fundamental fields over the dual coordinates is preserved. We impose that the generalized shift vector and the generalized lapse function satisfy ${\cal N}_{M}=0$ and $N=1$ respectively, while the generalized metric and the generalized dilaton satisfy a cosmological ansatz according to ${\cal H}_{MN}=A_{MN}(\tilde t,t)$ and $d=d(\tilde t, t)$. This framework preserves the duality group $O(1,1)\times O(D-1,D-1)$ and allows us to rewrite string cosmology equations in a duality invariant language as shown in (\ref{eqdft}). Interestingly enough, the matter contributions do not affect all the components of the generalized Einstein equation, and the DFT cosmologies can be directly related to string cosmologies upon field redefinitions of $\tilde p$ and $\tilde e$ according to (\ref{redef1}) and (\ref{redef2}) when $b_{i j}=0$. In a more general case, the present construction (\ref{eqdft}) allows one to obtain a family of duality invariant b--field contributions to string cosmology as in (\ref{Cosmosugra}). The results of this work pave the way for the understanding of the inclusion of statistical matter in DFT with applications to ST.  

We finish this section with a list of possible follow-up projects: 
\begin{itemize}
\item{\bf Duality invariant distribution function:} Making use of the construction given in \cite{phaseDFT} and considering a generalization of the Maxwell-Juttner distribution function, the generalized energy--momentum tensor for the perfect fluid might be computed from a kinetic approach. This is a broad line of investigation since the duality invariant distribution function would allow the inclusion of thermodynamic concepts in the double geometry such as the generalized entropy current or the study of equilibrium states.

\item{\bf Beyond perfect fluids in DFT:} The inclusion of a generalized entropy current from the phase space formulation should be the next step to interpret duality transformations on the generalized energy--momentum tensor as imperfect terms. At the level of the low energy limit of string theory, this issue was studied in \cite{GasVene}.

\item {\bf $\alpha'$-corrections:} A systematic way to introduce these corrections in a DFT background was given in \cite{gbdr}. The corrections are related to higher-derivative terms which deform the generalized Einstein tensor when matter terms are neglected. It would be interesting to test this procedure when matter deformations are taken into account in the double geometry. This might bring new perspectives for string cosmology.

\end{itemize}

\subsection*{Acknowledgements}
E.L thanks D. Marques for interesting discussions. E.L and N.M.G thank J.H. Park for e--mail correspondence and comments in the early stages of the work. N.M.G thanks CONICET for support.


\begin{thebibliography}{}
\bibitem{Siegel}
W.~Siegel, `Two vierbein formalism for string inspired axionic gravity', Phys.\ Rev.\ D {\bf 47} (1993) 5453, [hep-th/9302036]. 

W.~Siegel, `Superspace duality in low--energy superstrings', Phys.\ Rev.\ D {\bf 48} (1993) 2826, [hep-th/9305073].

W. ~Siegel, `Manifest duality in low--energy superstrings', In *Berkeley 1993, Proceedings, Strings '93* 353--363, and State U. New York Stony Brook - ITP-SB-93-050 11 p. (315661), [hep-th/9308133].

\bibitem{DFT}
C.~Hull and B.~Zwiebach, `Double Field Theory', JHEP {\bf 09} (2009) 099, [hep-th/0904.4664].

O.~Hohm, C.~Hull and B.~Zwiebach,
`Generalized metric formulation of Double Field Theory',
JHEP {\bf 08} (2010) 008,
[hep-th/1006.4823].

J. H. Park, `Comments on double field theory and diffeomorphisms', JHEP {\bf 06}
(2013) 098, [hep-th/1304.5946].

D. S. Berman, M. Cederwall, and M. J. Perry, `Global aspects of double geometry',
JHEP {\bf 09} (2014) 066, [hep-th/1401.1311].

\bibitem{DFTkorea}
I.~Jeon, K.~Lee and J.~H.~Park,
`Stringy differential geometry, beyond Riemann',
Phys.\ Rev.\ D {\bf 84} (2011) 044022,
[hep-th/1105.6294].

I.~Jeon, K.~Lee and J.~H.~Park, `Differential geometry with a projection: Application to Double Field Theory',
JHEP {\bf 04} (2011) 014,
[hep-th/1011.1324].

\bibitem{ReviewDFT1}
G.~Aldazabal, D.~Marques and C.~Nu\~nez,
`Double Field Theory: A Pedagogical Review',
Class.\ Quantum\ Grav.\  {\bf 30} (2013) 163001, [hep-th/1305.1907].

\bibitem{ReviewDFT2}
O.~Hohm, D.~Lust and B.~Zwiebach,
`The Spacetime of Double Field Theory: Review, Remarks, and Outlook',
Fortsch.\ Phys.\  {\bf 61} (2013) 926,  [hep-th/1309.2977].

D.~S.~Berman and D.~C.~Thompson,
`Duality Symmetric String and M--Theory',
Phys.\ Rept.\  {\bf 566} (2014) 1, [hep-th/1306.2643].

\bibitem{Gasperini}
M. Gasperini, `Elements of
string cosmology', Cambridge University Press, 2007, ISBN:978-0-521-18798-5.

\bibitem{SGC}
R. H. Brandenberger, C. Vafa, ``
Superstrings in the Early Universe'', Nucl. Phys. B 316 (1989) 391-410, 

\bibitem{VafaTseytlin}
A. A. Tseytlin and C. Vafa, `Elements of String Cosmology', Nucl. Phys. B 372 443-466 (1992)

\bibitem{90s}
K.A. Meissner l and G. Veneziano, `Symmetries of cosmological superstring vacua', Physics Letters B 267 33-36 (1991) 


\bibitem{SFD}
G. Veneziano, `Scale factor duality for classical and quantum strings', Physics Letters B 265 287-294 (1991) 

\bibitem{GasVene}
M. Gasperini and G. Veneziano, `O(d, d) covariant string cosmology', Phys. Lett. B 277
(1992) 256 [hep-th/9112044].

\bibitem{HZbullet}
O. Hohm and B. Zwiebach, `Duality invariant cosmology
to all orders in $\alpha'$', Phys. Rev. D 100, 126011 (2019),
[hep-th/1905.06963].

\bibitem{ParkCosmo}
S. Angus, K. Cho and J. H. Park, `Einstein Double Field Equations', Eur. Phys. J. C {\bf 78} (2018) 500, [hep-th/1804.00964].

S. Angus, K. Cho, G. Franzmann, S. Mukohyama and J.--H. Park, `$O(D,D)$ completion of
the Friedmann equations', Eur. Phys. J. C \textbf{80} (2020) 830, [hep-th/1905.03620].

\bibitem{Wu}
H. Wu and H. Yang, `Double Field Theory Inspired Cosmology', JCAP {\bf 07} (2014) 024, [hep-th/1307.0159].

\bibitem{Branden}
R. Brandenberger, R. Costa, G. Franzmann and A. Weltman, `T--dual cosmological
solutions in double field theory', Phys. Rev. D {\bf 99} (2019) 023531, [hep-th/1809.03482].

H. Bernardo, R. Brandenberger and G. Franzmann, `T--dual cosmological solutions in
double field theory II', Phys. Rev. D {\bf 99} (2019) 063521, [hep-th/1901.01209].

Y. Liu, `Dilatonic effect in double field theory cosmology', Gen Relativ Gravit \textbf{54}, 18 (2022) [hep-th/2112.15082]

\bibitem{Brandenmatter}
H. Bernardo, R. Brandenberger, G. Franzmann, `$O(d,d)$ covariant string cosmology to all orders in $\alpha'$', JHEP 02 (2020) 178, [hep-th/1911.00088]. 

\bibitem{alphapcosmo}
H. Bernardo and G. Franzmann, `$\alpha'$--cosmology: solutions and stability analysis', JHEP \textbf{05} (2020) 073 [hep-th/2002.09856]

\bibitem{Chouha}
H. Bernardo, P. R. Chouha, and G. Franzmann, `Kalb-Ramond backgrounds in $\alpha$'--complete cosmology', JHEP \textbf{09} (2021) 109 [hep-th/2104.15131]

\bibitem{gbdr}
W.~H.~Baron, E.~Lescano and D.~Marques,
``The generalized Bergshoeff-de Roo identification'',
JHEP {\bf 1811}, 160 (2018), [hep-th/1810.01427].

W. Baron, D. Marques, `The generalized Bergshoeff-de Roo identification. Part II', JHEP 01 (2021) 171, [hep-th/2009.07291]

E. Lescano,`$\alpha'$--corrections and their double formulation', [hep-th/2108.12246] 


\bibitem{phaseDFT}
E. Lescano and N. Mirón--Granese, `On the phase space in Double Field Theory', JHEP 07 (2020) 239, [hep-th/2003.09588].


\bibitem{Prk}
K. Lee and J. H. Park, `Covariant action for a string in ``doubled yet gauged” spacetime', Nucl. Phys.
B 880 (2014) 134 [hep-th/1307.8377].


\bibitem{RiemannDFT}
O.~Hohm and B.~Zwiebach, `On the Riemann Tensor in Double Field Theory', JHEP {\bf 05} (2012) 126, [hep-th/1112.5296].

\bibitem{Naseer}
U. Naseer, `Canonical formulation and conserved
charges of double field theory', JHEP {\bf 1510} (2015) 158, [hep-th/1508.00844].


\bibitem{Porrati}
Amit Giveon, Massimo Porrati, Eliezer Rabinovici, `Target space duality in string theory', Phys.Rept. 244 (1994) 77-202, [hep-th/9401139].

\bibitem{Susy}
I.~Jeon, K.~Lee and J.~H.~Park,
`Incorporation of fermions into double field theory,'
JHEP {\bf 1111} (2011) 025, [hep-th/1109.2035].

I.~Jeon, K.~Lee and J.~H.~Park, `Supersymmetric Double Field Theory: Stringy Reformulation of Supergravity', Phys.\ Rev.\ D {\bf 85} (2012) 081501 Erratum: [Phys.\ Rev.\ D {\bf 86} (2012) 089903], [hep-th/1112.0069].

O. Hohm and S. K. Kwak, `N=1 Supersymmetric Double Field Theory,' JHEP \textbf{1203} (2012) 080, [hep-th/1111.7293].

E. Lescano and A. Rodr\'iguez, `N = 1 Supersymmetric Double Field Theory
and the generalized Kerr--Schild Ansatz', JHEP {\bf 10} (2020) 148, [hep-th/2002.07751].

\bibitem{RR}
O. Hohm, S. K. Kwak, and B. Zwiebach, `Unification of Type II Strings and T--duality',
Phys. Rev. Lett. {\bf 107} (2011) 171603, [hep-th/1106.5452].

O. Hohm, S. K. Kwak, and B. Zwiebach, `Double Field Theory of Type II Strings', JHEP
{\bf 1109} (2011) 013, [hep-th/1107.0008].

I. Jeon, K. Lee, and J.--H. Park, `Ramond-Ramond Cohomology and O(D,D) T--duality',
JHEP {\bf 1209} (2012) 079, [hep-th/1206.3478].



\bibitem{scalarfluidgr}
V. Faraoni, `Correspondence between a scalar field and an effective perfect fluid', Phys. Rev. D {\bf 85} (2012) 024040, [gr-qc/1201.1448].

\bibitem{Cadabra} 
K.~Peeters, ``Introducing Cadabra: A Symbolic computer algebra system for field theory problems,''
[hep-th/0701238].








\end{thebibliography}
\end{document}